\documentclass[twocolumn,nofootinbib,prl,aps,superscriptaddress,tightenlines,preprintnumbers]{revtex4}

\usepackage{amsmath, amsfonts, amsthm, amssymb}
\usepackage{graphicx} 
\usepackage{cancel}
\usepackage{ytableau}
\usepackage{multirow}
\usepackage[normalem]{ulem}
\RequirePackage[colorlinks=true,urlcolor=blue,anchorcolor=blue,citecolor=blue,filecolor=blue,
              linkcolor=blue,menucolor=blue,linktocpage=true,pdfproducer=medialab,pagebackref=true]{hyperref}
\hypersetup{
 colorlinks   = true, %Colours links instead of ugly boxes
 urlcolor     = mygreen, %Colour for external hyperlinks
 linkcolor    = blue, %Colour of internal links
 citecolor   = blue %Colour of citations
}
\usepackage[capitalise]{cleveref}
\crefname{section}{Sec.}{Secs.}
\newcommand{\nn}{\nonumber}

\newcommand{\Sp}{\text{Sp}}
\newcommand{\SU}{\text{SU}}
\newcommand{\SO}{\text{SO}}
\newcommand{\U}{\text{U}}
\usepackage{xcolor}
\definecolor{mygreen}{rgb}{0.0, 0.5, 0.0}
\definecolor{myorange}{rgb}{1., 0.54902, 0.}

\def\diag{\mbox{diag}\,}

\usepackage{orcidlink}
\newcommand{\UPQ}{\U(1)_{\text{PQ}}}

\begin{document}
\setcounter{secnumdepth}{2}
\title{
\textbf{A High-Quality Composite Pati--Salam Axion}
}
\author{Tony Gherghetta\orcidlink{0000-0002-8489-1116}}
\email{tgher@umn.edu}
\affiliation{School of Physics and Astronomy, University of Minnesota, Minneapolis, Minnesota 55455, USA}

\author{Hitoshi Murayama\orcidlink{0000-0001-5769-9471}}
\email{hitoshi@berkeley.edu, hitoshi.murayama@ipmu.jp}
\affiliation{Department of Physics, University of California, Berkeley, California 94720, USA}
\affiliation{Kavli Institute for the Physics and Mathematics of the
  Universe (WPI), University of Tokyo,
  Kashiwa 277-8583, Japan}
\affiliation{Ernest Orlando Lawrence Berkeley National Laboratory, Berkeley, California 94720, USA}

\author{Pablo Qu\'ilez\orcidlink{0000-0002-4327-2706}}
\email{pquilez@ucsd.edu}
\affiliation{\it Department of Physics University of California, San Diego, California, USA
}

\begin{abstract}
We present a composite QCD axion model where the Peccei--Quinn (PQ) symmetry emerges as a high-quality, accidental symmetry. The axion potential is only modified by eight-fermion, dimension 12 operators, which if present at the Planck scale, allow for axion dark matter from misalignment while solving the strong CP problem. The model is an $\SU(N_c)$ gauge theory with ten flavors where the Pati--Salam unified subgroup $\SO(6)\times \SO(4) \subset \SU(10)_L$ and $\Sp(10)\subset \SU(10)_R$ are weakly gauged. 
The dynamics breaks $\SU(10)_L\times \SU(10)_R \rightarrow \SU(10)_V$ and the weakly-gauged groups to $\U(3)\times \U(2) \supset \SU(3)_c \times \SU(2)_L \times \U(1)_Y$, 
with the QCD axion identified as one of the Nambu-Goldstone bosons.
This axion has a relatively large coupling to photons while a residual $\bar{\theta}_{\rm eff}$ may be just below the current limit on the neutron electric dipole moment. If the dimension 12 operators are present near the GUT scale, they can cause domain wall networks to decay, allowing for axion dark matter even for the post-inflationary scenario.
\end{abstract}
\maketitle

\section{Introduction}

The QCD axion provides a compelling solution to the strong CP problem. By spontaneously breaking the Peccei--Quinn (PQ) symmetry $\UPQ$~\cite{Peccei:1977hh}, a global symmetry which is anomalous under $\SU(3)_c$, the resulting pseudo Nambu-Goldstone boson (pNGB)~\cite{Weinberg:1977ma,Wilczek:1977pj} dynamically cancels the strong CP phase $\bar\theta$, elegantly solving the strong CP problem. Importantly, the global $\UPQ$ can only be explicitly broken by QCD dynamics, otherwise the axion potential minimum could be shifted, spoiling the cancellation of the strong CP phase. However, there are good reasons to believe that global symmetries do not exist in theories of quantum gravity and the Peccei--Quinn symmetry cannot be respected by Planck-scale effects \cite{Georgi:1981pu,Kamionkowski:1992mf,Holman:1992us,Kallosh:1995hi,Barr:1992qq,Ghigna:1992iv,Alonso:2017avz,Alvey:2020nyh}.  
More recently the argument has been refined in the context of string theory ({\it e.g.}\/, \cite{Harlow:2018tng}). If taken seriously, the only exact symmetries of nature are gauge symmetries while global symmetries are violated by quantum gravity effects. Therefore, maintaining a high-quality, global $\UPQ$ symmetry in the presence of gravity is a necessary requirement for any QCD axion solution. 

Although such an axion may arise as one of the multiple axions generically resulting from string theory compactifications~\cite{Svrcek:2006yi,Arvanitaki:2009fg,Gavela:2023tzu}, in this paper we pursue an alternative approach to obtain a high-quality axion simply based upon conventional quantum field theory. 
A straightforward way to address the axion quality problem is to assume that the $\UPQ$ global symmetry is an {\it accidental}\/ symmetry. This can arise in composite axion models \cite{Kim:1984pt,Choi:1985cb} where the underlying strong dynamics plays the dual role of spontaneously breaking the global symmetry and naturally generating the scale of the axion decay constant. There are several composite axion models in the literature which solve (or alleviate) the axion quality problem by generating an accidental PQ symmetry. 
Originally this was achieved with several additional gauge groups \cite{Randall:1992ut,Redi:2016esr,Lillard:2018fdt}, 
but more recently theories with a simple confining chiral group have also been proposed \cite{Gavela:2018paw,Ardu:2020qmo,Contino:2021ayn}, which can also be understood geometrically in dual holographic descriptions~\cite{Cox:2019rro,Cox:2023dou}. 
In many cases, the underlying dynamics is not well-understood and therefore it is not clear whether the mechanisms generating the accidental symmetry are robust.

Instead, we consider QCD-like dynamics that robustly gives rise to a high-quality composite axion. We introduce a vector-like 
$\SU(N_c)$ confining group with ten flavors that admits an $\SU(10)_L \times \SU(10)_R$ flavor symmetry. The ten quarks transforming as $\bf N_c$ are also charged under a weakly-gauged $\SO(6)\times \SO(4) \subset \SU(10)_L$, while the ten anti-quarks transforming as ${\bf \bar{N}_c}$ are charged under a weakly-gauged $\Sp(10) \subset \SU(10)_R$. When the QCD-like $\SU(N_c)$ gauge theory confines, the usual chiral symmetry breaking breaks the flavor symmetry to its diagonal subgroup $\SU(10)_V$.
As a result, the quark condensates induce a dynamical breaking of $\SO(6) \times \SO(4) \times \Sp(10)$ to the diagonal $\U(3) \times \U(2) \simeq SU(3)_c \times U(1)_{B-L} \times SU(2)_L \times U(1)_{I_{3R}}$ gauge group, which is then further broken to the Standard Model gauge group $\SU(3)_c\times \SU(2)_L \times \U(1)_Y$ by a separate Higgs field.

A consequence of the $\SU(N_c)$ quark content is that there are two non-anomalous \U(1) global symmetries with respect to $\SU(N_c)$. One of these global symmetries has a QCD anomaly and therefore can be identified with the Peccei--Quinn symmetry. When the quark bilinear condensates form, they spontaneously break the PQ symmetry, leading to a pNGB that is identified with the QCD axion.
Furthermore, due to the gauge and Lorentz structure, there are no $\UPQ$-violating operators of dimension $< 12$ that modify the axion potential. The first PQ-violating operator contributing to the axion potential is a dimension twelve, eight-fermion operator. If this is generated at the reduced Planck scale $M_{P}=2.4 \times 10^{18}$~GeV, then the axion can still solve the strong CP problem provided the axion decay constant $f_a \lesssim 10^{11}$~GeV. Thus, as a result of the $\SU(N_c)$ strong dynamics, the axion quality problem is automatically solved. 

A unique feature of our model is that the Standard Model gauge group emerges from the strong dynamics associated with the Peccei--Quinn mechanism. This requires a particular embedding of the Standard Model fermions above the PQ scale. Given that $\SO(6)\times \SO(4) = \SU(4) \times \SU(2)_L \times \SU(2)_R$ is nothing but the Pati--Salam group, we can embed the Standard Model fermions into the usual ({\bf 4,2,1}) and ($\bf\bar{4}$,{\bf 1,2}) Pati--Salam representations above the PQ scale. Furthermore, since the $\SU(N_c)$ quarks are chiral with respect to $\SO(6)\times \SO(4)$ and $\Sp(10)$, the automatic PQ symmetry can also be understood as due to the chiral transformations of the $\SU(N_c)$ quarks under the weakly-gauged product gauge groups. Thus, since the $\SU(N_c)$ quarks remain vector-like under the $\SU(N_c)$ confining group, the underlying strong dynamics robustly predicts a high-quality, composite Pati--Salam axion which solves the strong CP problem. 
In fact, the embedding of the Standard Model into the Pati-Salam group, fixes the electromagnetic charges of the $\SU(N_c)$ fermions. This leads to a unique value for the anomaly ratio, $E/N=-7/3$ and therefore our model predicts a relatively large value for the axion-photon coupling that is much more accessible in future experiments than most models in the literature.

Moreover, our composite axion model is consistent with grand unification at a scale $\sim 10^{16}$~GeV. Axions in grand unified theories (GUTs) have
also been considered in previous works~\cite{Wise:1981ry,DiLuzio:2018gqe,Ernst:2018bib,
FileviezPerez:2019fku,FileviezPerez:2019ssf,Agrawal:2022lsp} and where the axion quality problem was solved with extra flavor gauge symmetries \cite{DiLuzio:2025jhv,DiLuzio:2020qio,DiLuzio:2020xgc}.
To achieve unification in our Pati-Salam axion model above the axion decay constant scale, in addition to the SM Higgs, we simply require an extra complex scalar field transforming as $({\bf 10,1,3})$ under the Pati-Salam group (which is part of the $\bf 126$ in $\SO(10)$) to remain below the GUT scale. This field is also responsible for achieving the final breaking of $\U(1)$ symmetries to $\U(1)_Y$. The $\SU(4) \times \SU(2)_L \times \SU(2)_R$ gauge couplings are found to unify into a single $\SO(10)$ coupling at a scale $M_{\rm GUT}\simeq 2\times 10^{16}$~GeV, for an axion decay constant scale $f_a\simeq 5\times 10^{11}$~GeV. Not only is this consistent with axion dark matter via the misalignment mechanism \cite{Preskill:1982cy,Abbott:1982af,Dine:1982ah}, but can also lead to an observable proton decay at the Hyper-Kamionkande experiment that is currently under construction (see, {\it e.g.}\/, \cite{Hisano:2022qll,Smy:2023miu}). 

Finally, the perturbative unification of the gauge couplings means that the axion potential in our composite axion model is not only screened from additional sources of CP-violation that arise from higher-dimensional operators generated at the GUT scale~\cite{Bedi:2022qrd}, but also from small instantons that could misalign the axion potential~\cite{Dine:1986bg,Csaki:2023ziz}. The unique, local gauge group structure of our model also makes it compatible with constraints from cosmology. In particular, the domain wall problem may be avoided because one of the $D=12$ PQ-violating operators explicitly breaks all the global discrete symmetry. 
This breaking is irrelevant for the case when the PQ-breaking occurs before or during inflation since domain walls can be inflated away, although at the cost of a low Hubble scale ($\lesssim 10^{6}$~GeV) during inflation to avoid large isocurvature fluctuations.
However, if the PQ breaking occurs after inflation, then the domain wall network can decay provided $f_a\sim 10^9$~GeV, producing axions as dark matter \cite{Kawasaki:2014sqa}. In this post-inflation scenario, any magnetic monopoles that form can be made to annihilate via the Langacker--Pi mechanism~\cite{Langacker:1980kd}, while tensor modes can now be detected in the CMB B-mode polarization, since the Hubble scale is no longer restricted by the isocurvature fluctuation.

\section{The Axion Quality Problem} % (fold)
\label{sec:PQquality}

The Peccei--Quinn symmetry is clearly not a gauge symmetry because it is anomalous under $\SU(3)_c$, by definition, to solve the strong CP problem. It is susceptible to being explicitly broken and hence we can hope that $\UPQ$ is an accidental symmetry due to gauge symmetries in the theory. For example, $\U(1)_B$ in the Standard Model is an accidental symmetry because there are no gauge-invariant operators that can violate $\U(1)_B$ for dimension $D<6$ (see Ref.~\cite{Grinstein:2024jqt} for a recent systematic analysis of accidental symmetries).

There is a more closely related example. In three-flavor massless QCD, there is an $\SU(3)_L \times \SU(3)_R \times \U(1)_B$ global symmetry, which is dynamically broken to $\SU(3)_V \times \U(1)_B$ by the quark bilinear condensate $\langle \bar{q}_i q_j \rangle = \Lambda_{\rm QCD}^3 \delta_{ij}$. The electromagnetic $\U(1)_{\rm EM}$ is embedded into $\SU(3)_V$ as the generator $\frac{1}{3} {\rm diag}(2, -1, -1)$. On the other hand, $\pi^0$ is a Nambu--Goldstone boson for the $I_3 = \frac{1}{2}{\rm diag}(1, -1, 0)$ generator of dynamically broken $\SU(3)_A$, an accidental global symmetry. The $\U(1)$ symmetry associated with $I_3$ is anomalous under $\U(1)_{\rm EM}$ which gives rise to a $\pi^0 F \tilde{F}$ coupling. However, in the dimension $D=3$ quark mass terms $m_u \bar{u} u + m_d \bar{d} d + m_s \bar{s} s$, $I_3$ is explicitly broken by the up and down masses
and hence $\pi^0$ acquires a significant correction to its potential, namely its mass squared $m_{\pi}^2 \simeq (m_u + m_d) \Lambda_{\rm QCD}^3 / f_\pi^2$. To obtain the QCD axion based on a similar accidental symmetry from new strong dynamics but avoiding a significant correction to the axion potential, the corresponding exotic quark mass term must be forbidden and hence the theory must be a chiral gauge theory.

If the $\UPQ$ breaking is due to an elementary complex scalar field $\phi$, there may well be explicit $\UPQ$-violating operators suppressed by the Planck scale \cite{Kamionkowski:1992mf}
\begin{align}
    {\cal L} \supset \frac{1}{M_{P}^{D-4}} (c_{\cancel{\rm PQ}} \phi^{D} + c^*_{\cancel{\rm PQ}} \phi^{*D})\,,
    \label{eq:U1PQLag}
\end{align}
where $D\geq 5$ is an integer and $c_{\cancel{\rm PQ}}\equiv |c_{\cancel{\rm PQ}}| e^{i\delta}$ is an arbitrary complex coefficient with phase $\delta$.
The operators in \cref{eq:U1PQLag} can shift the axion potential minimum away from where the CP-violating $\bar\theta$-parameter is canceled. Assuming both $|c_{\cancel{\rm PQ}}|$ and $\delta$ are $O(1)$, an approximate upper bound on $f_a$ for operators with dimension $D=(9,10,11,12)$ is
\begin{align}
    f_a & \lesssim (10^{8.5},10^{9.5},10^{10.3},10^{10.9})\, ,
\end{align}
for $\delta=1$. Thus, to avoid astrophysical limits $f_a \gtrsim 4 \times 10^8$~GeV, we need $D\geq 9$. To obtain axion dark matter based on the misalignment mechanism, the axion decay constant must satisfy $f_a \gtrsim 10^{11}$~GeV, requiring $D\geq 11$. 

\begin{figure}[t]
\includegraphics[scale=0.5]{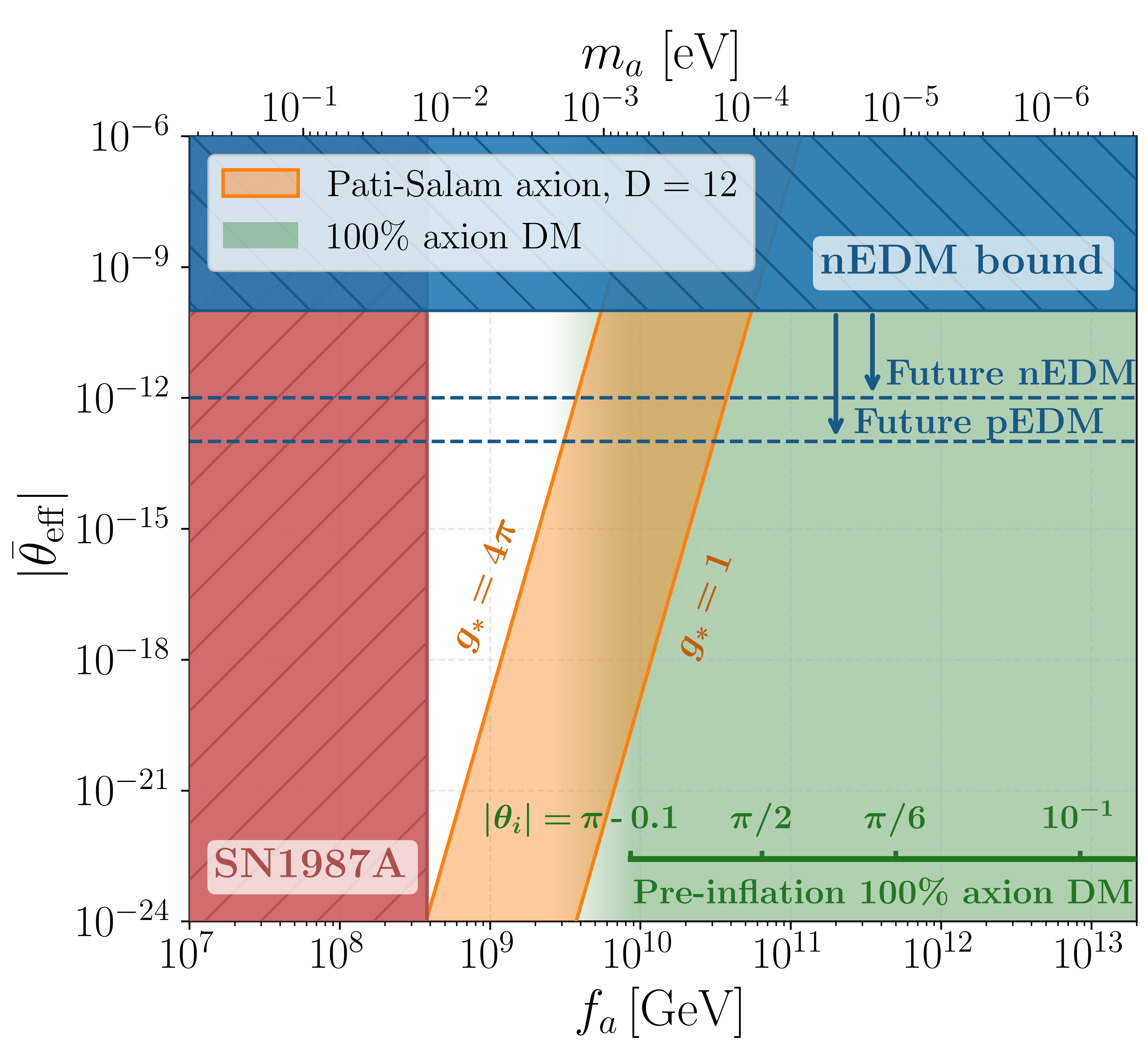}
\caption{The expected contribution to $\bar{\theta}_{\rm eff}$ for the Pati--Salam axion (arising from the operator in \cref{eq:PQV} for $N_c=4$ and $|c_{\cancel{PQ}}|\sin\delta=0.1$) is shown in {\color{myorange}\textbf{orange}} as a function of $f_a$. The region where the Pati--Salam axion can account for $100\%$ of DM in the pre-inflationary scenario is shown in green, along with the required initial misalignment angle $|\theta_i|\equiv| a_i/f_a|$. For $f_a\lesssim10^{10}$GeV the green color fades away because successful dark matter requires increasingly fine-tuned initial misalignment angles $\theta_i\sim \pi$. Exclusion limits from SN1987A and nEDM experiments are shown as hatched regions, while projected sensitivities of future nEDM and pEDM experiments are indicated with dashed lines. The SN1987A bound is stronger than the model-independent limit in 
\cite{Springmann:2024ret}, since in our hadronic axion model, axion–nucleon interactions dominate supernova cooling \cite{Carenza:2019pxu}.
}
\label{Fig:PatiSalamQuality}
\end{figure}

In a composite axion model, the order parameter of $\UPQ$ breaking is not an elementary scalar but a composite operator, and thus we can hope that the lowest dimension operator that can modify the axion potential necessarily has a large dimension $D\geq 12$. As we will see below, axion dark matter  can be realized in our scenario within the constraint on $\bar{\theta}_{\rm eff}$, as shown in Fig.~\ref{Fig:PatiSalamQuality}.

\section{The Pati--Salam Axion model}
\label{sec:SO6model}

We now present a simple, composite model of the QCD axion based on QCD-like dynamics that simultaneously addresses the axion quality problem and dynamically breaks a product Pati-Salam gauge group containing the Standard Model.

We consider a vector-like $\SU(N_c)$ theory with ten flavors 
that admits an $\SU(10)_R \times \SU(10)_L \times \U(1)_B$ flavor symmetry. Within the ten quark flavors $Q$ in ${\bf N_c}$, we gauge $\SO(6) \times \SO(4) \subset \SU(10)_L $, while for the ten anti-quarks $P$ in ${\bf \bar{N}_c}$, we gauge $\Sp(10) \subset \SU(10)_R$ (see Table~\ref{tab:SO6}). In this way, overall the theory is a chiral gauge theory that forbids $D=3$ mass terms for the $Q,P$ quarks, but remains vector-like with respect to the $\SU(N_c)$ strong dynamics. Moreover, for the $\Sp$ gauge groups to be free from the Witten anomaly~\cite{Witten:1982fp}, the strong gauge group must have even $N_c$.
\begin{table}[thb]
\centering
\begin{tabular}{|c||c|c|c|c||c|c|} \hline
\multirow{2}{*}{} & \multicolumn{4}{|c||}{$G_{\rm local}$} & \multicolumn{2}{c|}{$G_{\rm global}$} \\ \cline{2-7}
& $\SU(N_c)$  & $\Sp(10)$ &  $\SO(6)$ & $\SO(4)$  & $\U(1)_B$ & $\UPQ$  \\\hline \hline
$Q_6$ & ${\bf N_c}$ & {\bf 1} & {\bf 6} &  {\bf 1}  & $+1$ & $+2$ \\ \hline
$Q_4$ & ${\bf N_c}$ & {\bf 1} & {\bf 1} & {\bf 4} &  $+1$ & $-3$ \\ \hline \hline
$P$ & $\bf\bar{N}_c$ &  {\bf 10}  & {\bf 1} & {\bf 1} 
& $-1$ & $0$ \\ \hline 
\end{tabular}
\caption{The fermion content of the $\SU(N_c)$ theory with ten flavors in the Pati--Salam composite axion model under local and global symmetries. All fermions are left-handed Weyl fermions. The global $\U(1)$ symmetries are anomaly-free under the strong $\SU(N_c)$ gauge group but not under $\SO(6)\times\SO(4)\times\Sp(10)$ and hence are accidental.
}
\label{tab:SO6}
\end{table}

The Standard Model gauge group is embedded into the Pati--Salam group $\SO(6)\times \SO(4) \cong \SU(4) \times \SU(2)_L \times \SU(2)_R$. Its $({\bf 4,2,1})$ representation contains the $\SU(2)_L$-doublet quarks, $Q$, and leptons, $L$, while $({\bf \bar{4},1,2})$ contains the $\SU(2)_L$-singlet fermions, ${\bar u}, {\bar d}, {\bar e}$ and the right-handed neutrino $N$.

There are two accidental $\U(1)$ symmetries that are anomaly-free under the strong $\SU(N_c)$ gauge group. One of them is identified with $\UPQ$ because it is anomalous under $\SU(3)_c$ embedded into the Pati--Salam gauge group. Note that the $\SU(N_c)$ quarks are charged under the PQ symmetry (see Table~\ref{tab:SO6}), while the Standard Model fermions do not have PQ charges and therefore our model is similar to hadronic or KSVZ axion models~\cite{Kim:1979if,Shifman:1979if,Zhitnitsky:1980tq}.

The usual chiral symmetry breaking in QCD-like theories have meson-like order parameters
\begin{align}
%    {\cal M}^r_i = 
    \langle P^r Q_i \rangle \approx \Lambda^3 \delta^r_i,
    \label{eq:condensate}
\end{align}
where $\Lambda$ is the dynamical scale associated with $\SU(N_c)$ and $i$ ($r$) refers to the $\SU(10)_L$ ($\SU(10)_R$)  index. The condensate \cref{eq:condensate}
breaks the $\SU(10)_L \times \SU(10)_R$ flavor symmetry to its diagonal subgroup $\SU(10)_V$ and as a result, the $\SO(6) \times \SO(4) \times \Sp(10)$ gauge group to the diagonal $\U(3) \times \U(2)$.  The unbroken $\U(3)$ is $\SU(3)_c \times \U(1)_{B-L}$, while $\U(2)$ is $\SU(2)_L \times \U(1)_{I_{3R}}$. The $\U(1)_{B-L} \times \U(1)_{I_{3R}}$ is then further broken by an additional Higgs field to $\U(1)_Y$, where the Standard Model hypercharge is given by $Y = \frac{1}{2}(B-L)+ I_{3R}$.
A successful chiral symmetry breaking is possible for $N_f \lesssim 3 N_c$ (see \cite{Kondo:2021osz} and references therein) and we most likely need $N_c \geq 4$. 

Importantly, the bilinear fermion condensates in \cref{eq:condensate} also spontaneously break the $\UPQ$ symmetry, leading to a massless pNGB, identified as the QCD axion. On the other hand, $\U(1)_B$ remains unbroken, as expected for the vector-like $\SU(N_c)$ strong dynamics~\cite{Vafa:1983tf}.

There are ${\rm dim}[\SU(10)_L\times \SU(10)_R/SU(10)_V]=99$ pNGBs from the fermion bilinear condensates. On the other hand, the number of NGBs eaten by the gauge bosons is ${\rm dim}\big[(\SO(6)\times \SO(4) \times \Sp(10)) / (\U(3) \times \U(2))\big] =63$, and therefore $99-63=36$ pNGBs remain.
These transform under the $\U(3)\times \U(2)$ unbroken gauge group as $({\bf 8_0,1_0})\oplus ({\bf 1_0,3_0}) \oplus ({\bf 3_1+{\bar 3}_{-1},2_1+2_{-1}}) \oplus ({\bf 1_0,1_0})$. 
All non-singlet pNGBs acquire positive mass-squared from the unbroken gauge interactions $m_{\rm pNGB}^2 \sim g^2 f_{\rm PQ}^2$ (analogously to the $\pi^+-\pi^0$ electromagnetic mass difference \cite{Bardeen:1988zw}), where $g$ denotes the gauge coupling and $f_{\rm PQ}$ the PQ breaking scale.
The only singlet is the QCD axion and while it is massless at this stage, it will eventually obtain the usual mass via the QCD anomaly.

\section{PQ-violating Operators}
\label{sec:PQVops}

The most important point is that this model does not allow for any $\UPQ$-violating higher-dimension operators below $D=12$ that would misalign the axion potential.
This can be seen by noting that to build invariants under $\SO(6)$, we need the following fermion bilinears 
\begin{align}
\label{eq:Q6s}
    (Q_6^2)^{\{a,b\}}&\equiv\delta^{ij} (Q_{6i\alpha}^{a} \epsilon^{\alpha\beta}Q_{6j\beta}^{b}), \\
    (Q_6^2)^{[a,b]}_{\mu\nu}&\equiv\delta^{ij} 
    (Q_{6i\alpha}^{a} \sigma_{\mu\nu}^{\alpha\beta} Q_{6j\beta}^{b}),
    \label{eq:Q6a}
\end{align}
where $a,b$ are $\SU(N_c)$ indices, $i,j$ are $\SO(6)$ indices, $\alpha,\beta$ are spinor indices. The notation $\{a,b\}$ indicates that two $\SU(N_c)$ indices are symmetric, while $[a,b]$ implies antisymmetric indices. The case for $Q_4$ is completely analogous. On the other hand, the invariants under $\Sp(10)$ are
\begin{align}
\label{eq:Pa}
    (P^2)_{[a,b]} &\equiv J_{rs} (P^{r\alpha}_{a} \epsilon_{\alpha\beta}P^{s\beta}_{b})\,, \\
    (P^2)_{\{a,b\}}^{\mu\nu} &\equiv J_{rs} 
    (P^{r\alpha}_{\{a} \sigma^{\mu\nu}_{\alpha\beta} P^{s\beta}_{b\}})\,,
    \label{eq:Ps}
\end{align}
where $r,s$ are $\Sp(10)$ indices and $J_{rs}$ is antisymmetric. Note that the symmetry properties under the interchange $a\leftrightarrow b$ are opposite for $Q_6$ and $P$. It is clear that there are no four-fermion operators that are Lorentz- and gauge-invariant. Exactly the same is true for $Q_4$ and $P$. 

Therefore, the lowest-dimension operator that acquires an expectation value and hence is an order parameter of symmetry breaking is
\begin{align}
    {\cal O}_{12} \equiv \frac{1}{M_{P}^8}
    (Q_{4,6}^2)^{\{a,b\}} (Q_{4,6}^2)^{[c,d]}_{\mu\nu}
    (P^2)_{[c,d]} (P^2)_{\{a,b\}}^{\mu\nu}\, ,
    \label{eq:PQV}
\end{align}
which has dimension $D=12$ and nonzero PQ charge. Note that there are other dimension 12 operators that can be formed from different combinations of Eqs.~\eqref{eq:Q6s},\eqref{eq:Q6a},\eqref{eq:Pa} and \eqref{eq:Ps}, although not all of these operators may be independent.
The operator \cref{eq:PQV} can be rearranged using the Fierz transformation 
as
\begin{align}
    J_{rs} J_{tu} \delta^{ij} \delta^{kl} (P^r Q_i) (P^s Q_k) (P^t Q_j) (P^u Q_l)\,,
\end{align}
so that it is totally symmetric under the interchange of the order parameter $(PQ)$. 
This operator does obtain an expectation value while violating $\UPQ$ and hence gives a correction to the axion potential. As a result, the axion almost cancels the $\bar\theta$ parameter leaving a residual $\bar\theta_{\rm eff}$ that can be estimated to be
\begin{align}
  |\bar\theta_{\rm eff}|\simeq \frac{|c_{\cancel{\rm PQ}} \sin\delta|\,  g_*^{10}}{4!4!M_{\rm P}^8}\left(\frac{4N_c}{\sqrt{13}}\right)^{12}\frac{2}{N_c}\frac{f_a^{12}}{\chi_{QCD}}\,,
  \label{Eq:thetaeff}
\end{align}
where $\chi_{QCD}\equiv{m_\pi^2 f_\pi^2} \frac{{m_u\,m_d}}{(m_u+m_d)^2}=m_a^2f_a^2$ is the QCD topological susceptibility,\footnote{As is customary, the axion decay constant is redefined as  $f_a \equiv f_{\rm PQ}/N$  to absorb the color anomaly factor $N = 4N_c$, see \cref{Eq:AxionEff,Eq:anoamliesPQandprime}.} $c_{\cancel{\rm PQ}}$ is the overall coefficient of the operator in \cref{eq:PQV} and $\delta$ is the phase mismatch between this operator and $\bar\theta$. 
While the relation between the confinement scale $\Lambda$ and the PQ-breaking scale $f_{\rm PQ}$ can be estimated using naive dimensional analysis~\cite{Manohar:1983md,Cohen:1997rt,Gavela:2016bzc}, we consider the more general relation $\Lambda\simeq g_* f_{\rm PQ}$, 
where $g_*$ represents the typical coupling between the composite bound states satisfying $1 \lesssim g_* \lesssim 4 \pi$. 
The resulting residual $\bar\theta_{\rm eff}$ obtained from \cref{Eq:thetaeff} remains below the current limits $10^{-10}$ as long as $f_a \lesssim 10^{11}$~GeV (see Fig.~\ref{Fig:PatiSalamQuality}). This condition allows the composite axion to be the dark matter with an order one initial misalignment angle.

Note that there can be operators of lower dimensions that violate $\UPQ$, however they do not modify the axion potential. For example, for the $\SU(N_c=4)$ strong gauge group, the \textit{baryonic}\footnote{Here, \textit{baryonic} refers to fully antisymmetric contractions of $N_c$ fundamentals with an $\epsilon$ tensor. For even $N_c$, such operators only yield  bosonic composites, in contrast to the familiar fermionic baryons of QCD.} four-fermion operators $(Q_4)^4$ and $(Q_6)^4$ with dimension $D=6$ would be possible. Both of them have $\SO(4)$- and $\SO(6)$-singlet combinations, which may exist in the Lagrangian at the Planck scale 
\begin{equation}
\frac{1}{M_{P}^2} (c_4 (Q_4)^4_{\bf 1}+c_6 (Q_6)^4_{\bf 1})\,,
\label{eq:baryonop}
\end{equation}
where $c_{4,6}$ are complex coefficients.
Besides the PQ charge, these operators have a nonzero baryonic charge and therefore they do not acquire expectation values solely from the strong dynamics since the QCD-like dynamics cannot break baryon number \cite{Vafa:1983tf}. 
As also noticed in Ref.~\cite{Contino:2021ayn}, only PQ-violating operators with vanishing vectorial charge can form vacuum condensates and contribute to the axion potential, which is the case for the operators in \cref{eq:PQV}.

However, after confinement the operators in \cref{eq:baryonop} do generate linear potential terms for the (scalar) baryon fields, which can induce baryon expectation values. For instance,
below the strong scale $\Lambda$, the scalar baryons $(Q_4)^4_{\bf 1}\approx \Lambda^5 {\cal B}_4 e^{12 i a/f_a}$ and $(Q_6)^4_{\bf 1}\approx \Lambda^5 {\cal B}_6 e^{-8 i a/f_a}$ have a potential 
\begin{align}
    \Delta V &\approx \Lambda^2 |{\cal B}_4|^2 + \Lambda^2 |{\cal B}_6|^2 \nonumber \\
    & - \frac{1}{M_{P}^2} \Lambda^5 (c_4 e^{12 i a/f_a}{\cal B}_4 + c_6 e^{-8 i a/f_a}{\cal B}_6) + h.c.,
    \label{eq:baryonV}
\end{align}
which leads to the expectation values 
\begin{align}
    \langle {\cal B}_4\rangle = \frac{1}{M_{P}^2} c_4^* \Lambda^3 e^{-12 i a/f_a}, \quad
     \langle{\cal B}_6 \rangle= \frac{1}{M_{P}^2} c_6^* \Lambda^3 e^{8 i a/f_a}.
    \label{eq:B}
\end{align}
Inserting \cref{eq:B} back into the potential \cref{eq:baryonV}, we obtain
\begin{align}
    \Delta V \approx -(c_4^* c_4 +c_6^* c_6) \frac{\Lambda^8}{M_{P}^4} \ ,
\end{align}
and the axion dependence cancels. However, the axion dependence in the potential may arise if there exists an operator that contains both $Q_4$ and $Q_6$. For instance, if $\SO(4) \times \SO(6)$ is embedded into $\SO(10)$ (see ~\cref{sec:SO10}), it may generate an operator
\begin{align}
    \lefteqn{
    \frac{1}{M_{\rm GUT}^8} (\bar{Q}_4 \sigma^\mu Q_6 \bar{Q}_4 \sigma_\mu Q_6)^2 
    } \nonumber \\
    & \approx \frac{\Lambda^{10}}{M_{\rm GUT}^8} {\cal B}_4^* {\cal B}_6
    \approx \frac{\Lambda^{16}}{M_{\rm GUT}^8 M_{P}^4} c_4 c_6^* e^{-4 i a/f_a},
\end{align}
where $M_{\rm GUT}$ is the grand unified theory (GUT) scale. Due to the large suppression, this operator is clearly harmless. On the other hand for the $\SU(N_c=6)$ strong gauge group, baryon operators have no $\SO(4)$ or $\SO(6)$ singlet combinations (see~\cref{app:baryonops}). Therefore, there are no gauge-invariant Planck-scale operators of this type and consequently no contributions to the axion potential.

Furthermore, our model is not sensitive to contributions from ``small instantons'' \cite{Dine:1986bg,rubakov:1997vp,Gherghetta:2016fhp,Agrawal:2017ksf,Gaillard:2018xgk,Bedi:2022qrd,Csaki:2023ziz} either. In particular, the $\SU(N_c)$ instantons do not generate a potential for the axion because $\UPQ$ is non-anomalous under $\SU(N_c)$. The $\Sp(10)$ instantons generate a $P^{N_c}$ operator, which does not have a $\UPQ$ charge (however they can make the baryon unstable, see \cref{sec:cosmology}). The $\SO(4)$ instantons generate a
{$\frac{1}{\Lambda^{26}} (Q_4)^{8} F^{12} e^{-2\pi/\alpha_{\SO(4)} (\Lambda)}$}operator where $F$ are Standard Model fermions. Even if the $F$ are all contracted with %some scalars, 
Higgs fields, using 
{$\alpha^{-1}_{\SO(4)} (\Lambda) \approx 40$} (see Fig.~\ref{Fig:GrandUnificationplot} below), it is even more suppressed than the Planck-scale $D=12$ operator. The $\SO(6)$ instantons are similarly harmless. Thus, in our model the dominant contribution that can misalign the axion potential is due to the dimension 12 operators \cref{eq:PQV}.

The axion dark matter from the misalignment mechanism prefers $f_a \gtrsim 10^{10}$~GeV while the operator \cref{eq:PQV} may shift the axion potential minimum to induce a residual $\bar\theta \sim 10^{-10}$, near the current upper limit (see Fig.~\ref{Fig:PatiSalamQuality}). The resultant neutron electric dipole moment (nEDM) may be detectable. The current best limit $d_n < 1.8 \times 10^{-26}~e\cdot$cm is from PSI UltraCold Neutron (UCN) experiment. On the other hand, the TUCAN experiment at TRIUMF is aiming at $10^{-27}~e\cdot$cm \cite{TUCAN:2022koi} while there are proposals to reach the sensitivity even down to $10^{-28}~e\cdot$~cm at Grenoble \cite{Balashov:2007zj} or \href{https://j-parc.jp/researcher/Hadron/en/pac_1001/pdf/KEK_J-PARC-PAC2009-11.pdf}{J-PARC} (see Ref.~\cite{EuropeanStrategyforParticlePhysicsPreparatoryGroup:2019qin} for details on short/mid-term planned sensitivities).
Even though nEDM experiments currently provide the leading bounds on $\bar \theta$, storage ring
facilities~\cite{Alexander:2022rmq} are expected to provide limits on the proton EDM (pEDM) of the order of $d_p\sim 10^{-29}\,e\cdot$cm in the near future~\cite{Balashov:2007zj}. Therefore, the projections of the pEDM bounds are competitive with those
for future nEDM  measurements, as shown in \cref{Fig:PatiSalamQuality}.

\section{Axion-photon coupling}

In hadronic axion models (like KSVZ and composite axion models), the SM quarks lack PQ charges, so the axion-photon coupling depends on freely assignable electromagnetic charges of exotic fermions. This differs from DFSZ, where SM quarks and leptons contribute to anomalies, fixing their charges and restricting possible axion-photon couplings. Our model is a mixture: while SM quarks remain PQ-neutral, the axion-photon coupling is still fixed because the massless fermions have specific transformation properties under the Pati--Salam gauge group, fixing their electromagnetic charges via the Standard Model embedding in the unified group.

\ytableausetup{boxsize=0.5em}
Therefore, our model is completely predictive. We can compute the anomaly factors $E,N$ that determine the axion coupling to photons and gluons
\begin{align}
 \mathcal{L}_{a}&\supset\frac{\alpha_{EM}}{8\pi}E\frac{a}{f_{\rm PQ}} F_{\mu\nu} \tilde F^{\mu\nu} + \frac{\alpha_s}{8\pi} N\frac{a}{f_{\rm PQ}}G_{a\,\mu\nu} \tilde G_a^{\mu\nu} \nonumber\\ 
 &=\frac{\alpha_{EM}}{8\pi}\frac{E}{N}\frac{a}{f_{a}} F_{\mu\nu} \tilde F^{\mu\nu} + \frac{\alpha_s}{8\pi} \frac{a}{f_a}G_{a\,\mu\nu} \tilde G_a^{\mu\nu}\,,
 \label{Eq:AxionEff}
 \end{align}
where $\alpha_{EM} (\alpha_{s})$ is the electromagnetic (QCD) fine structure constant. In the second line of \cref{Eq:AxionEff} we have defined $f_a\equiv f_{\rm PQ}/N$ and the anomaly factors are
$E=2 \sum_f q_{{\rm PQ},f} \, q_{{\rm EM},f}^2; N= 2\sum_f q_{{\rm PQ},f}\,  T(R_{{\rm QCD},f})$, where $q_{{\rm PQ},f}$ $(q_{{\rm EM},f})$ is the PQ (electric) charge of the fermion $f$,   
%is the electric charge 
and $T(R_{{ \rm QCD},f})$ is the Dynkin index of the corresponding $\SU(3)_c$ representation, defined as $T(R) \delta_{a b}\equiv \operatorname{Tr}\left(t_R^a t_R^b\right)$, with 
$T(\ydiagram{1})=1/2$. The anomaly coefficients are calculated to be 
\begin{align}
&\UPQ\times \left[\, \SU(3)_c\right]^2: \quad N=2\cdot N_c\cdot (+2)=4N_c\,, \nn\\ 
&\UPQ\times \left[\,\U(1)_{\rm EM}\right]^2:\nn\\
&E=2 N_c\Big[ (+2)\cdot 6\cdot\Big(\frac{\pm1}{3}\Big)^2 +(-3)\cdot 2\cdot (\pm 1)^2\Big]=-\frac{28}{3}N_c\,,
\label{Eq:anoamliesPQandprime}
\end{align}
where further details of the embedding are given in \cref{app:SMembed}.
Combining the results in \cref{Eq:anoamliesPQandprime}, we finally obtain the ratio $E/N=-7/3$.
After taking into account the mixing of the axion with the QCD neutral mesons, the axion-photon coupling is modified to become \cite{diCortona:2015ldu}
\begin{align}
 g_{a \gamma \gamma}=\frac{\alpha_{\mathrm{EM}}}{2 \pi f_a}\left(-\frac{7}{3}-1.92(4)\right)\,.
 \label{eq:axionphoton}
\end{align}

\begin{figure}[t]
\centering
\includegraphics[width=0.49\textwidth]{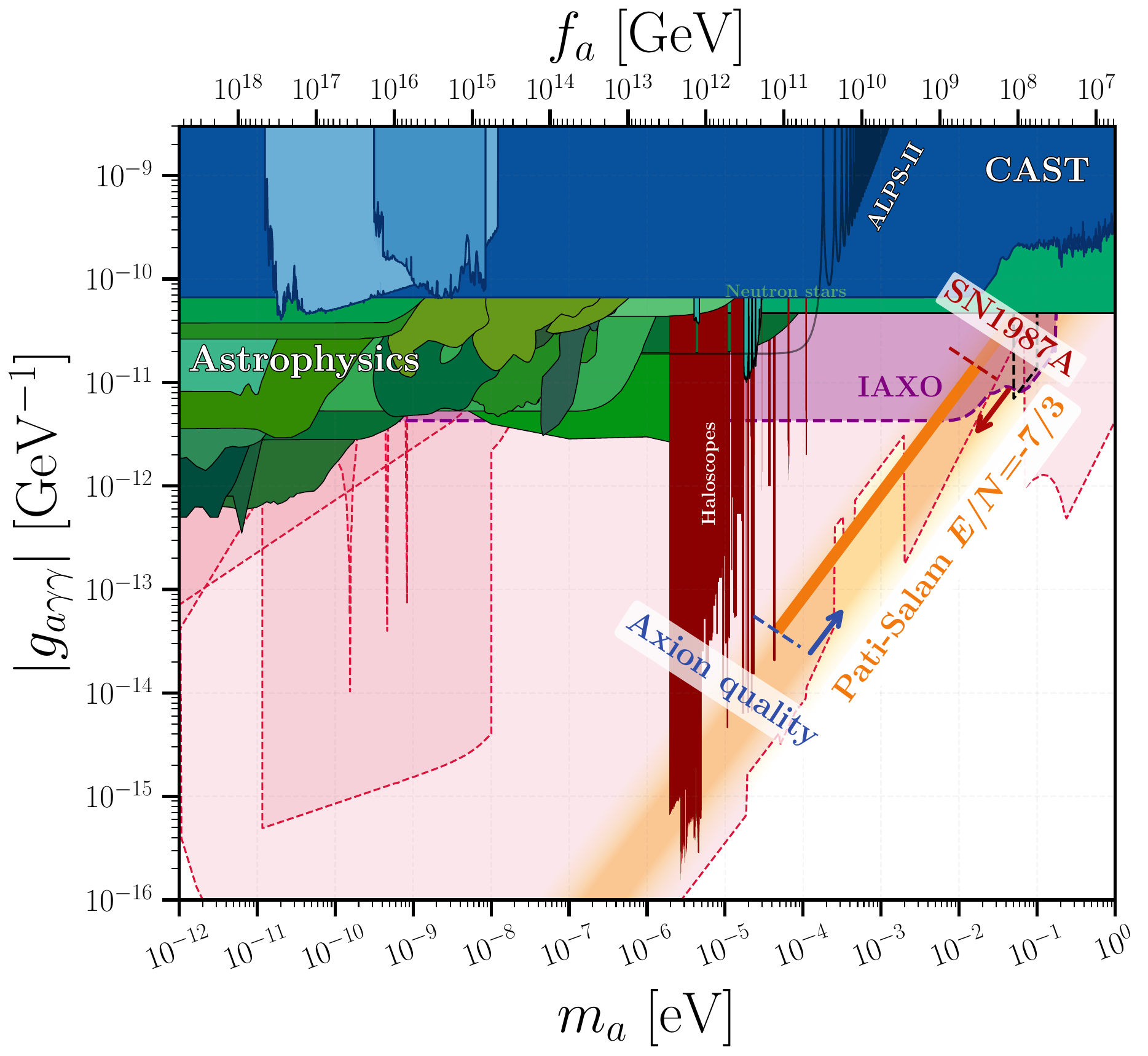}
\caption{The limits on the axion-photon coupling as a function of the axion mass adapted from Ref.~\cite{Axionlimits}. The prediction for the Pati--Salam axion is shown as a  {\color{myorange}\textbf{orange}}, thick line, which corresponds to $E/N=-7/3$. The range of allowed values of $f_a$ is bounded from below by the SN1987A bound and from above by imposing that the operator in \cref{eq:PQV} does not spoil the axion quality.
\label{Fig:PatiSalamPhotons}}
\end{figure}

Since the ratio $E/N$ is negative for our Pati-Salam axion, \cref{eq:axionphoton} implies a rather large value for $|g_{a \gamma \gamma}|$. This larger axion-photon coupling lies at the upper end of the axion band and thus will be more accessible in future experiments. 
As shown in \cref{Fig:PatiSalamPhotons}, even IAXO could test our model for $m_a\gtrsim 4~\text{meV}$.
%$m_a\gtrsim 4\times 10^{-3}\text{eV}$. 
This value should be compared with the typical $E/N$ ratios predicted by various axion models, see \textit{e.g.}~\cite{DiLuzio:2016sbl,DiLuzio:2017pfr,Plakkot:2021xyx,DiLuzio:2024xnt}. For instance, in both DFSZ models and grand unified theories with a simple unifying group, one finds $E/N = 8/3$ \cite{Wise:1981ry,Agrawal:2022lsp}, which leads to a significantly smaller value of $g_{a\gamma\gamma}$ compared to our model. Similarly, twelve out of the fifteen \textit{preferred} KSVZ axion models with a single exotic quark, as studied in Ref.~\cite{DiLuzio:2016sbl}, predict values of $E/N$ that yield a smaller axion-photon coupling compared to the value obtained in our Pati–Salam composite axion model.

\section{$\SO(10)$ Grand Unification}
\label{sec:SO10}

An interesting feature of our Pati--Salam axion model is that the Pati--Salam group can be unified into $\SO(10)$ at a scale $2\times 10^{16}$~GeV, compatible with an axion decay constant scale needed for axion dark matter via the misalignment mechanism.

To achieve the unification we require two minimal assumptions. First, the usual Standard Model Higgs doublet is introduced as part of $H=({\bf 1,2,2})$ transforming under the Pati--Salam group and contained in the $\bf 10$ of $\SO(10)$. Second, we introduce an extra scalar field $\Phi$ in the Pati--Salam representation $({\bf 10,1,3})$ which belongs to the ${\bf 126}$ in $\SO(10)$. These split SO(10) multiplets are consistent with only allowing fields that obtain VEVs to appear at a particular scale or lower~\cite{delAguila:1980qag,Dimopoulos:1984ha}.
While we assume only one Standard Model Higgs doublet below $f_a$, it is also possible to have unification with two Higgs doublets at low energies~\cite{Djouadi:2022gws}.
Moreover, the scalar field $\Phi$ is also required to break $\U(1)_{B-L} \times \U(1)_{I_{3R}}\rightarrow \U(1)_Y$ by its vacuum expectation value $\langle \Phi \rangle$, and generate mass for right-handed neutrinos. For simplicity, we assume $\langle \Phi \rangle \approx f_a$ in our unification analysis. Note that thermal leptogenesis can work when the lightest, right-handed neutrino mass, $M_1 \gtrsim 10^9$~GeV and hence possible within our model \cite{Buchmuller:2004nz}.

The Standard Model gauge couplings are run up to the scale $f_a$. Due to the dynamical breaking to the diagonal gauge group at the scale $f_a$, the couplings are then matched onto the $\SO(6)\times \SO(4)\times \Sp(10)$ = $\SU(4)\times \SU(2)_L \times \SU(2)_R\times \Sp(10)$ couplings as follows
\begin{align}
    \label{eq:a1fa}
      \alpha_1^{-1}(f_a)&= \frac{2}{5}\alpha_{\SU(4)}^{-1}(f_a) +\frac{3}{5}\alpha_{\SU(2)_R}^{-1}(f_a)+\alpha_{\Sp(10)}^{-1}(f_a)\,,\\
      \label{eq:a2fa}
      \alpha_2^{-1}(f_a)&= \alpha_{\SU(2)_L}^{-1}(f_a)+\alpha_{\Sp(10)}^{-1}(f_a)\,,\\
       \label{eq:a3fa}
      \alpha_3^{-1}(f_a)&= \alpha_{\SU(4)}^{-1}(f_a)+\alpha_{\Sp(10)}^{-1}(f_a)\,,    
\end{align}
where $\alpha_{1,2,3}^{-1}$ are the Standard Model couplings associated with $\U(1)_Y,\SU(2)_L, \SU(3)_c$, respectively. Above the scale $f_a$ the couplings run with the following $\beta$-function coefficients
\begin{align}   &\left(b_{\SU(4)},b_{\SU(2)_L},b_{\SU(2)_R},b_{\Sp(10)}\right)\nonumber\\&=\frac{1}{3}\left(2N_c -23,2N_c-9,2N_c+11,N_c-66\right)\,,
%\left(-\frac{7}{3},\frac{7}{3},9,-\frac{58}{3}\right)\,,
    \label{eq:bcoeff}
\end{align}
where the renormalization group equation (RGE) is given by $d\alpha_i^{-1}(\mu)/d\ln{\mu}=-b_i/(2\pi)$ and the matter content contains three generations of Standard Model fermions and the two complex scalar fields $H$ and $\Phi$. The solution to the RGEs with $\beta$-function coefficients \cref{eq:bcoeff} and matching conditions Eqs.~\eqref{eq:a1fa}-\eqref{eq:a3fa} fixes the scale of unification and the axion decay constant.  The couplings unify at a scale $M_{\rm GUT}\simeq 2\times 10^{16}$~GeV with a unified $\SO(10)$ coupling value $\alpha_{\rm GUT}^{-1} \simeq 41.9-0.85N_c$ and an axion decay constant scale $f_a \simeq 5\times 10^{11}$~GeV.  The running of the couplings is shown in Fig.~\ref{Fig:GrandUnificationplot} for $N_c=8$ where $\alpha_{\rm GUT}^{-1} \simeq 35$. Note that the unification scales are computed using only the one-loop running. By including threshold effects and two-loop running the unification could also accommodate a lower\footnote{It is also possible to have an even lower value of $f_a\simeq 10^9$~GeV, as required for a post-inflation scenario, by introducing additional split SO(10) multiplets.} $f_a \sim 10^{11}$~GeV and thus be compatible with the orange region shown in Fig~\ref{Fig:PatiSalamQuality}. Similarly, once threshold corrections are included, the proton decay $p \rightarrow e^+ \pi^0$ may be observable at the Hyper-Kamiokande experiment \cite{Smy:2023miu} slated to start taking data in 2027.

Finally, note that the $\SO(10)$ mass relations are incompatible with the experimentally observed fermion masses and mixings with just one $\bf 10$ and $\bf 126$. This can be remedied by adding an extra $\bf 10$ at the GUT scale~\cite{Babu:1992ia,Bajc:2005zf} (which mildly affects the one-loop unification in \cref{Fig:GrandUnificationplot}), altering the mass relations via mixing the Standard Model generations with extra ${\bf 16} + {\bf\bar{16}}$ multiplets or introducing additional operators at the GUT scale.
It would be interesting to perform a fit as considered in Ref.~\cite{Ohlsson:2019sja}.

\section{Cosmology}
\label{sec:cosmology}

A common cosmological challenge in axion models, particularly for composite axions, is the domain wall problem. While the axion potential breaks the continuous $U(1)_{\rm PQ}$ symmetry explicitly, it often preserves a discrete subgroup. When this residual symmetry is spontaneously broken as the axion field acquires a vacuum expectation value, topologically stable domain walls form. These domain walls dilute slowly and, if sufficiently long-lived, can eventually dominate the energy density of the Universe, thereby disrupting standard cosmological evolution. This issue is especially severe in models of accidental composite axions (see e.g.~\cite{Kim:1984pt,Lu:2023ayc}), where the confining dynamics typically yield a domain wall number $N_{\rm DW} > 1$. Moreover, the same gauge symmetries that render the $U(1)_{\rm PQ}$ symmetry accidental often protect the residual discrete symmetry, thereby preventing the rapid decay of all domain walls. As a result, such models are generally only viable in a pre-inflationary cosmological scenario.
\begin{figure}[t]
\centering
\includegraphics[width=0.49\textwidth]{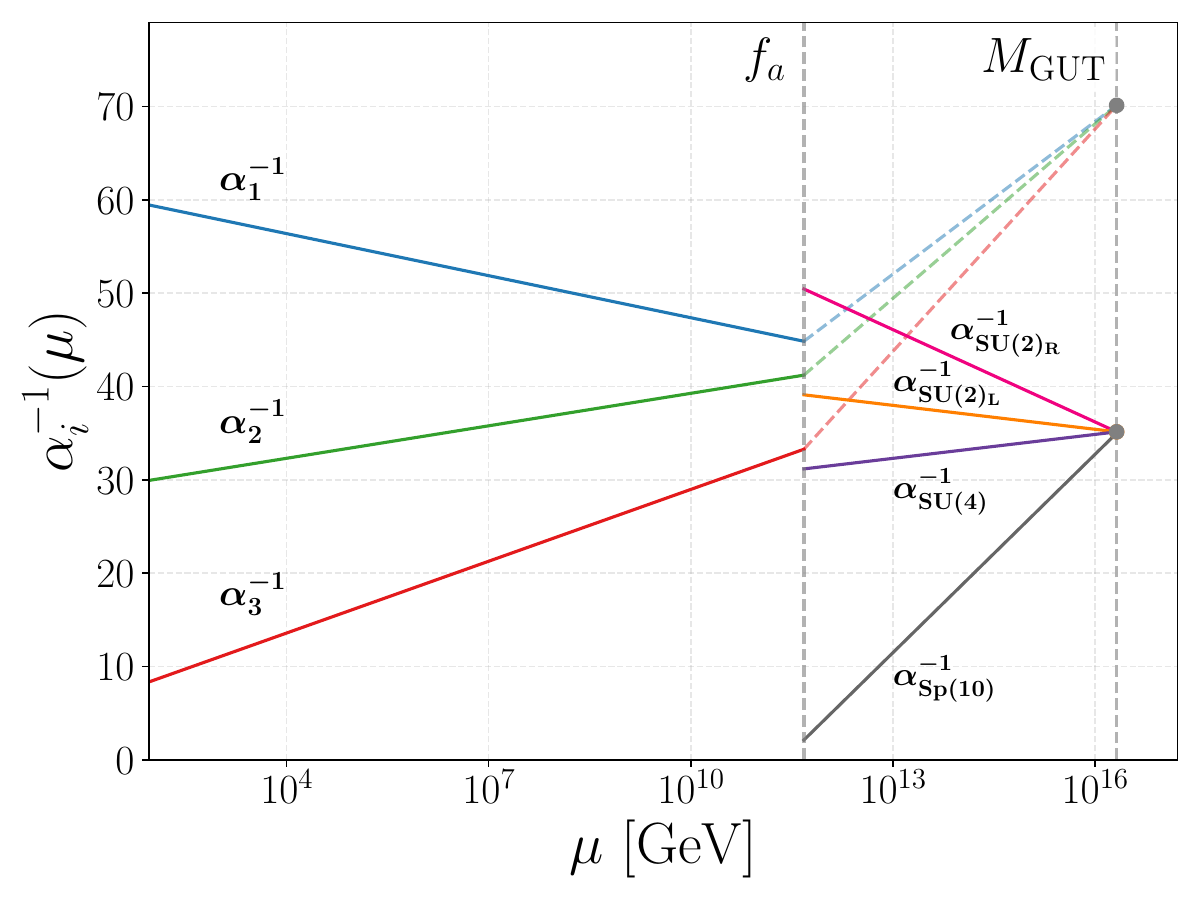}
\caption{The running of the Standard Model gauge couplings using the one-loop renormalization group equations and assuming $N_c = 8$. Note that above the scale $f_a\simeq 5\times 10^{11}$~GeV, the evolution of the individual $\SU(4)\times \SU(2)_L \times \SU(2)_R\times \Sp(10)$ gauge couplings is shown with solid lines that meet at the GUT scale $M_{\rm GUT}\simeq 2\times 10^{16}$~GeV. The dashed lines, in contrast, correspond to the linear combinations defined in Eqs.~\eqref{eq:a1fa}-\eqref{eq:a3fa}, depicting the continuous evolution of the Standard Model couplings.}
\label{Fig:GrandUnificationplot}
\end{figure}

While our model does work well in a pre-inflation scenario, we can potentially avoid the domain wall problem even when the PQ breaking occurs after inflation.
This can be seen by considering the anomalies under $\SU(N_c)$ and $\SU(3)_c$ while ignoring the anomalies under the other weakly-coupled groups. Then the ${\mathbb Z}_{4N_c}$ subgroup of $\UPQ$ is non-anomalous,
\begin{align}
	Q_6 &\rightarrow e^{2\pi i (+2) k /4 N_c} Q_6,  \quad
	Q_4 \rightarrow e^{2\pi i (-3) k /4 N_c} Q_4, \nonumber \\
	P & \rightarrow P \qquad
	(k = 0, 1, \cdots, 4N_c-1).
    \label{eq:discrete}
\end{align}
The VEVs \cref{eq:condensate} spontaneously break the ${\mathbb Z}_{4N_c}$ discrete symmetry leading to stable domain walls. If the PQ-breaking occurs before the end of inflation, the domain walls are inflated away and the problem does not exist. The cosmological abundance of axions must then be due to the misalignment mechanism. The axion decay constant can be as large as $f_a \approx 10^{11}$~GeV while maintaining the axion quality even when the PQ-violating $D=12$ operators 
%$\frac{1}{M_{P}^8}{\cal O}_{12}$ 
\eqref{eq:PQV} are present. 

We can expect the axion to be an excellent dark matter candidate. There is a constraint, however, that the Hubble expansion rate $H_I$ towards the end of the inflation needs to be sufficiently small, $H_{I} \lesssim 10^{-5} f_a \approx 10^{6}$~GeV, to avoid excessive isocurvature fluctuations. In this case, we do not expect tensor modes to be observable in the CMB B-mode polarization.

However, the PQ breaking after inflation is also possible under some conditions. If the PQ-violating operator 
\begin{align}
    {\cal O}_{12} \equiv \frac{1}{M_{\cancel{\rm PQ}}^8}
    (Q_{4}^2)^{\{a,b\}} (Q_{6}^2)^{[c,d]}_{\mu\nu}
    (P^2)_{[c,d]} (P^2)_{\{a,b\}}^{\mu\nu}\, ,
    \label{Eq:PQVDW}
\end{align}
is present at the scale $M_{\cancel{\rm PQ}}$ below the Planck scale $M_{P}$, the discrete symmetry \cref{eq:discrete} is explicitly broken to ${\mathbb Z}_{2}$ for $k=2N_c$ which is nothing but the center of $\SO(4)$. 
Therefore, due to this lifting of the vacuum degeneracy, domain walls move and collapse against each other, and the remaining walls bounded by the ${\mathbb Z}_2$ string~\cite{Lazarides:1982tw}, shrink and disappear. 
Note that the collapse may not be sufficiently efficient when the center is large \cite{Lu:2023ayc}, but in our case, since the center is ${\mathbb Z}_{2}$, this problem most likely does not exist.  

While the operator \cref{Eq:PQVDW} will destabilize the domain walls and cause them to decay, successful cosmology also requires that this happens sufficiently fast and that the emitted axions account for the observed dark matter relic density. Estimating the axion relic density from domain wall decay is non-trivial and demands numerical simulations to assess the viability of this scenario \cite{Kawasaki:2014sqa,Gorghetto:2020qws}. It is crucial to evaluate the compatibility of a relatively large PQ-violating operator \cref{Eq:PQVDW}—needed for fast domain wall decay—with a sufficiently small effective $\bar{\theta}$ to satisfy nEDM constraints. 
Some estimates~\cite{Kawasaki:2014sqa} indicate that $f_a$ can be as low as $4\times10^8-10^9$~GeV to account for the dark matter, if there is a percent-level fine-tuning in the phase $\delta$ and $M_{\cancel{\rm PQ}}\sim 10^{15}-10^{17}$GeV for our operator~\cref{Eq:PQVDW}. 
This lower value of $f_a$ makes the axion more accessible to experimental searches.

The problem with magnetic monopoles can be avoided as well. There are two $\U(1)$ gauge groups, $\U(1)_{B-L}$ and $\U(1)_{I_{3R}}$ after the Pati--Salam breaking and hence there are two types of magnetic monopoles. We assume that they are broken to $\U(1)_Y$ by the ({\bf 10,1,3}) Higgs field discussed in \cref{sec:SO10}. Many monopoles annihilate by the resultant magnetic flux tubes, but some combinations may remain. However they can all annihilate via the Langacker--Pi mechanism~\cite{Langacker:1980kd}, where $\U(1)_{Y}$ is temporarily broken at some point in the early Universe and then restored at a later time. 
If this happens, the PQ breaking can then occur after inflation, which in turn allows for a large inflation scale and the corresponding tensor mode signal in the CMB B-mode.

The formation of $\SU(N_c)$ baryons leads to another potentially dangerous relic.
%Another potential dangerous relic is $\SU(N_c)$ baryons. 
However as shown in \cref{Fig:GrandUnificationplot}, $\Sp(10)$ becomes quite strong at the axion decay constant scale and hence instanton effects can make the baryon $P^{N_c}$ unstable against decays to pNGBs. Given the chiral symmetry breaking, $P$ and $Q$ form Dirac fermions with the constituent mass $\sim f_{\rm PQ}$ and hence there is no distinction between $Q^{N_c}$ and $\bar{P}^{N_c}$. A more accurate estimate of its lifetime depends on the assumption about the gauge unification and is beyond the scope of this paper.

\section{Conclusion}

We have presented a composite axion model that achieves a high-quality, Peccei--Quinn symmetry as an accidental symmetry of a well-understood $\SU(N_c)$ strong dynamics. By weakly gauging an $\SO(6)\times \SO(4)\times \Sp(10)$ subgroup of the $\SU(10)_L\times \SU(10)_R$ flavor symmetry, operators with small dimension can be forbidden. For $N_c\geq 4$, the lowest dimension operator that acquires a vacuum expectation value with nonzero PQ charge has dimension 12.  
This allows for an axion decay constant scale that is consistent with axion dark matter produced via the misalignment mechanism. In addition, the dimension 12 operator may lead to a non-zero $\bar\theta$ just below the current limit on the neutron electric dipole moment, as shown in Fig.~\ref{Fig:PatiSalamQuality}.

The $\SU(N_c)$ strong dynamics, dynamically breaks $\SO(6)\times \SO(4)\times \Sp(10)\rightarrow \U(3)\times \U(2)$, which after a further breaking of the $\U(1)$ symmetries (by an additional light, scalar field $\Phi \subset {\bf 126}$ of $\SO(10)$) to $\U(1)_Y$), gives rise to the Standard Model gauge group. Interestingly, the Standard Model gauge group emerges from the weakly-gauged part of the $\SU(N_c)$ flavor symmetry which also plays a crucial role in realizing the PQ global symmetry as an accidental symmetry. 

Given that $SO(6) \times SO(4)$ is equivalent to the Pati-Salam group $\SU(4) \times \SU(2)_L \times \SU(2)_R$, the Standard Model fermions can be embedded in the conventional way. This predicts a negative value for the anomaly ratio $E/N=-7/3$ and therefore gives rise to a relatively large axion-photon coupling that can be seen at the IAXO experiment, as shown in Fig.~\ref{Fig:PatiSalamPhotons}. Furthermore, the light scalar field $\Phi$, responsible for breaking the $\U(1)$ symmetries, also allows the $\SU(4)\times \SU(2)_L\times \SU(2)_R$ gauge couplings to unify at a scale $M_{\rm GUT}\simeq 2\times 10^{16}$~GeV into an $\SO(10)$ gauge group for $f_a\simeq 5\times 10^{11}$~GeV, as shown in Fig.~\ref{Fig:GrandUnificationplot}. This unification scale may lead to proton decay at the future Hyper-Kamionkande experiment. 

Our model is also compatible with constraints from early Universe cosmology. If the PQ breaking occurs before or during inflation then there is no domain wall problem, although the tensor mode signal in the CMB B-mode polarization becomes extremely suppressed. Interestingly, however, our model would also allow for the PQ-breaking to occur after inflation, and still avoid the domain wall problem. This is because the $D=12$ PQ-violating operator in \cref{Eq:PQVDW} explicitly breaks the discrete symmetry to ${\mathbb Z}_{2}$, which is just part of the local $\SO(4)$ symmetry. Dark matter is now possible with $f_a$ as low as $10^9$~GeV, monopoles can be made to annihilate and tensor modes can be detected in the CMB B-mode. This provides an example of a post-inflation composite axion model.

Thus, using the well-understood dynamics of $\SU(N_c)$ gauge theories, we have constructed a high-quality, Pati-Salam axion based on accidental symmetries that provides an intriguing and well-motivated target for future experiments.

\section*{Acknowledgements}
We thank Xiaochuan Lu, Wen Yin, Lawrence Hall, Mario Reig, and Keisuke Harigaya for useful discussions.
The work of T.\,G. is supported in part by the Department of Energy under Grant No. DE-SC0011842 at the University of Minnesota. The work of H.\,M.\ is supported by the Director, Office of Science, Office of High Energy Physics of the U.S. Department of Energy under the Contract No. DE-AC02-05CH11231, by the NSF grant PHY-2210390, by the JSPS Grant-in-Aid for Scientific Research JP23K03382, MEXT Grant-in-Aid for Transformative Research Areas (A) JP20H05850, JP20A203, Hamamatsu Photonics, K.K, and Tokyo Dome Corportation. In addition, H.\,M.\ is supported by the World Premier International Research Center Initiative (WPI) MEXT, Japan.
The work of P.\,Q. is supported by the U.S. Department of Energy under grant number DE-SC0009919 and partially supported by the European Union's Horizon 2020 research and innovation programme under the Marie Sk\l odowska-Curie grant agreement No. 101086085-ASYMMETRY.
Likewise, H.\,M., T.\,G. and P.\,Q. thank the CERN theory group for their warm hospitality during the \emph{Crossroads between Theory and Phenomenology} Program, where this work was initiated.

\appendix
\section{SM embedding in $\SO(6)\times \SO(4)\times \Sp(10)$}
\label{app:SMembed}

The SM gauge group $\SU(3)_c\times \SU(2)_L\times \U(1)_Y$ is embedded into the Pati--Salam group $\SO(6)\times \SO(4) = \SU(4) \times \SU(2)_L \times \SU(2)_R$ in the standard way.
The  quark and lepton $\SU(2)_L$ doublets, $Q$ and $L$, are unified into a single $({\bf 4,2,1})$ representation, while the $\SU(2)_L$ singlets $\bar u,\bar d, \bar e$, together with the right-handed neutrino $N$,  are contained in a $({\bf\bar 4,1,2})$. 

To compute the anomaly ratio $E/N$, we also require the relevant branching rules for $Q_6$ given by:
\begin{align}
{\bf 6}_{\SO(6)}  &\to {\bf 3_{-\frac{2}{3}}}+{\bf {\bar 3}_{\frac{2}{3}}}
%(3,-2/3)+(\bar 3,+2/3) 
\text{ under } %(\SU(3)_c,\U(1)_{B-L})
\SU(3)_c\times \U(1)_{B-L} \nn\\ 
    &\to {\bf 3_{-\frac{1}{3}}}+{\bf {\bar 3}_{\frac{1}{3}}}
    %(3,-1/3)+(\bar 3,+1/3) 
    \text{ under } %(\SU(3)_c,\U(1)_{Y}) 
    \SU(3)_c\times \U(1)_{Y} \nn\\ 
    &\to {\bf 3_{-\frac{1}{3}}}+{\bf {\bar 3}_{\frac{1}{3}}}
    %(3,-1/3)+(\bar 3,+1/3) 
    \text{ under } 
    %(\SU(3)_c,\U(1)_{\rm EM}) 
    \SU(3)_c\times\U(1)_{\rm EM}
    \nn
\end{align}
while the branching rules of $Q_4$ are
\begin{align}
% Q_4:  
{\bf 4}_{\SO(4)/\Sp(4)}  &\to {\bf 2_{\frac{1}{2}}}+{\bf 2_{-\frac{1}{2}}}
%(2,+1/2)+(2,-1/2) 
\text{ under } 
%(\SU(2)_L,\U(1)_{3R})
\SU(2)_L\times \U(1)_{3R}\nn\\ 
    &\to {\bf 2_{\frac{1}{2}}}+{\bf 2_{-\frac{1}{2}}}
    %(2,+1/2)+(2,-1/2) 
    \text{ under } 
    %(\SU(2)_L,\U(1)_{Y}) 
    \SU(2)_L\times \U(1)_{Y}\nn\\ 
    &\to ({\bf +1})+({\bf 0})+({\bf 0})+({\bf -1}) \text{ under } \U(1)_{\rm EM}\nn
\end{align}
The $P$ quarks also decompose in the idential manner.
The branching rules are obtained by using the embeddings of the different Abelian factors which correspond to
\begin{align}
  B-L&= \diag\left(\frac{1}{3},\frac{1}{3},\frac{1}{3},-1\right)\subset \SU(4)_{PS}\,,\nn \\ 
  I_{3L}&= \diag\left(\frac{1}{2},-\frac{1}{2}\right)\subset \SU(2)_L\,,\nn \\ 
  I_{3R}&= \diag\left(\frac{1}{2},-\frac{1}{2}\right)\subset \SU(2)_R\,,\nn \\ 
  Y&=I_{3R}+\frac{1}{2}(B-L)\,,\nn \\ 
  Q&=I_{3L}+Y\,.
\end{align}

\section{Baryon Operators}
\label{app:baryonops}

In this appendix, we present further details to understand the properties of the $\UPQ$-violating baryon operators, discussed in \cref{sec:PQVops},
under the gauge groups $\SO(6)\times \SO(4)$.

The baryon operators are built out of the Weyl fermions $(Q_{4,6})^a_{i\alpha}$, where $a, i , \alpha$ are the $\SU(N_c), \SO(n) (n=4~{\rm or}~6)$ and spinor indices, respectively. 
Since baryons have their colors contracted with $\epsilon$-tensors for $\SU(N_c)$ that antisymmetrize color indices, and fermion fields anticommute, the rest of the indices for $\SO(n)$ and spin must be overall totally symmetric. Considering the spin-flavor group $\SU(2n)$, the representation must be totally symmetric consisting of $N_c$ boxes. We only consider even $N_c = 2k$ for $k=1,2,\dots$ and use $N_c=2k=8$ for the illustrations below.
The baryons are in the following $\SU(2n)$ representation with the multiplicity
\ytableausetup{mathmode,  boxsize=1.2em}
\begin{align}
    \ydiagram{8}
    \qquad
    \binom{2n+2k-1}{2k}.
    \label{eq:Brep}
\end{align}
When decomposed under $\SU(n) \times \SU(2)$, corresponding to the flavor and spin symmetry, respectively, this representation is a product of identical Young tableaux with $k+j$ boxes in the first row and $k-j$ in the second. Therefore the decomposition is
\begin{align}
    \ydiagram{8} &\raisebox{0.4em}{\ $\otimes$\ } \ydiagram{8} \nonumber \\
    \raisebox{-0.2em}{\ $\oplus$\ } 
    \ydiagram{7,1} &\raisebox{-0.2em}{\ $\otimes$\ } \ydiagram{7,1} \nonumber \\
    \raisebox{-0.2em}{\ $\oplus$\ } 
    \ydiagram{6,2} &\raisebox{-0.2em}{\ $\otimes$\ }  \ydiagram{6,2} \nonumber \\
    \raisebox{-0.2em}{\ $\oplus$\ } 
    \ydiagram{5,3} &\raisebox{-0.2em}{\ $\otimes$\ }  \ydiagram{5,3} \nonumber \\
    \raisebox{-0.2em}{\ $\oplus$\ } 
    \ydiagram{4,4} &\raisebox{-0.2em}{\ $\otimes$\ }  \ydiagram{4,4}.
    \label{eq:baryontableaux}
\end{align}
Note that the reason why there are only two rows is because $\SU(2)$ has only two possible indices $\alpha=1,2$. 
The spin part has a multiplicity of $2j+1=9,7,5,3,1$, respectively, while the corresponding $\SU(n)$ Young tableaux has a multiplicity of 
\begin{align}
    N_{nkj}=\frac{(n+k+j-1)!(n+k-j-2)!(2j+1)}{(n-1)!(n-2)!(k+j+1)!(k-j)!} \ .
\end{align}
One can indeed verify that
\begin{align}
\sum_{j=0}^k %\frac{(n+k+j-1)!(n+k-j-2)!(2j+1)}{(n-1)!(n-2)!(k+j+1)!(k-j)!} 
(2j+1) N_{nkj}%\nonumber \\
=\binom{2n+2k-1}{2k}\,,
%    \lefteqn{\binom{2n+2k-1}{2k}=} \nonumber \\
%    &\sum_{j=0}^k \frac{(n+k+j-1)!(n+k-j-2)!(2j+1)}{(n-1)!(n-2)!(k+j+1)!(k-j)!} (2j+1)\,,
\end{align}
in agreement with the multiplicity in \cref{eq:Brep}.

The case $j=0$ in the last line of \cref{eq:baryontableaux} is a Lorentz scalar, a scalar baryon, and hence can be a term in the Lagrangian. Yet we need to see if this scalar contains an $\SO(n)$ singlet which can be obtained by contracting indices (boxes) with $\delta_{ij}$. Since $\delta_{ij}$ is symmetric, pairs of boxes in the same row can be contracted. When $k$ is even, all indices can be contracted and made an $\SO(n)$ singlet. On the other hand when $k$ is odd, one box in each row remains uncontracted and this pair cannot be contracted because their indices are antisymmetric.
%They cannot be contracted because their indices are antisymmetric. 
This is why there are $\SO(6)$- or $\SO(4)$-invariant scalar baryon operators when $N_c$ is a multiple of four, such as $\SU(4)$ or $\SU(8)$, while there are no such
%$\SO(6)$- or $\SO(4)$-invariant scalar baryon 
operators for $N_c = 2$~mod~$4$, such as $\SU(6)$. Similarly, it is easy to see that scalar baryon operators formed from the $P$ Weyl fermions can be $\Sp(10)$ invariant using $J_{ij}$ tensors that contract two boxes in each column for any even $N_c$.

\bibliographystyle{utphys}
\bibliography{BiblioNew}

\end{document}